\documentclass{vldb}

\usepackage{booktabs} 

\vldbTitle{Query2Vec: An Evaluation of NLP Techniques for Generalized Workload Analytics}
\vldbAuthors{Shrainik Jain, Bill Howe, Jiaqi Yan, and Thierry Cruanes}
\vldbDOI{https://doi.org/TBD}

\usepackage{graphicx}
\usepackage{balance}  
\usepackage{url}
\usepackage{float}
\usepackage{array}
\usepackage{caption}
\usepackage{subcaption}
\usepackage{amsmath,amssymb}
\usepackage{algorithm, algorithmic}
\usepackage[usenames, dvipsnames]{color}
\newcolumntype{L}[1]{>{\raggedright\let\newline\\\arraybackslash\hspace{0pt}}m{#1}}
\newcolumntype{C}[1]{>{\centering\let\newline\\\arraybackslash\hspace{0pt}}m{#1}}
\newcolumntype{R}[1]{>{\raggedleft\let\newline\\\arraybackslash\hspace{0pt}}m{#1}}

\newfloat{procedure}{htbp}{loa}
\floatname{procedure}{Procedure}

\begin{document}

\title{Query2Vec: An Evaluation of NLP Techniques for Generalized Workload Analytics}

\numberofauthors{4} 

\author{
%
%
\alignauthor
Shrainik Jain\\
       \affaddr{University of Washington}\\
       \email{\small{shrainik@cs.washington.edu}}\\
\alignauthor
Bill Howe\\
       \affaddr{University of Washington}\\
       \email{\small{billhowe@cs.washington.edu}}
\and
\alignauthor
Jiaqi Yan\\
       \affaddr{Snowflake Computing}\\
       \email{\small{jiaqi.yan@snowflake.net}}\\
\alignauthor
Thierry Cruanes\\
       \affaddr{Snowflake Computing}\\
       \email{\small{thierry.cruanes@snowflake.net}}\\
}

\maketitle
\begin{abstract}
We consider methods for learning vector representations of SQL queries to support generalized workload analytics tasks, including workload summarization for index selection and predicting queries that will trigger memory errors. We consider vector representations of both raw SQL text and optimized query plans, and evaluate these methods on synthetic and real SQL workloads.  We find that general algorithms based on vector representations can outperform existing approaches that rely on specialized features.  For index recommendation, we cluster the vector representations to compress large workloads with no loss in performance from the recommended index.  For error prediction, we train a classifier over learned vectors that can automatically relate subtle syntactic patterns with specific errors raised during query execution.  Surprisingly, we also find that these methods enable transfer learning, where a model trained on one SQL corpus can be applied to an unrelated corpus and still enable good performance. We find that these general approaches, when trained on a large corpus of SQL queries, provides a robust foundation for a variety of workload analysis tasks and database features, without requiring application-specific feature engineering.
\end{abstract}
\section{Introduction}




Extracting patterns from a query workload has been an important technique in database systems research for decades, used for a variety of tasks including workload compression~\cite{chaudhuri2002compressing}, index recommendation~\cite{chaudhuri:03}, modeling user and application behavior~\cite{tran:15,jain:16a,yu:92}, query recommendation~\cite{querie}, predicting cache performance~\cite{sapia:00,dan:95}, and designing benchmarks~\cite{yu:92}.

We see a need for generalized, automated techniques that can support all of these applications with a common framework, due to three trends: First, workload heterogeneity is increasing.  In loosely structured analytics environments (e.g., ``data lakes"), ad hoc queries over ad hoc datasets tend to dominate routine queries over engineered schemas~\cite{farid:16}, increasing heterogeneity and making heuristic-based pattern analysis more difficult.  Second, workload scale is increasing.  With the advent of cloud-hosted, multi-tenant databases like Snowflake which receive tens of millions of queries every day, database administrators can no longer rely on manual inspection and intuition to identify query patterns queries. 
Third, new use cases for workload analysis are emerging.  User productivity enhancements, for example SQL debugging~\cite{grust:13} and database forensics~\cite{pavlou:13}, are emerging, motivating a more automated analysis of user behavior patterns. 

\begin{figure}[t]
\centering
\includegraphics[keepaspectratio=true,scale=0.9]{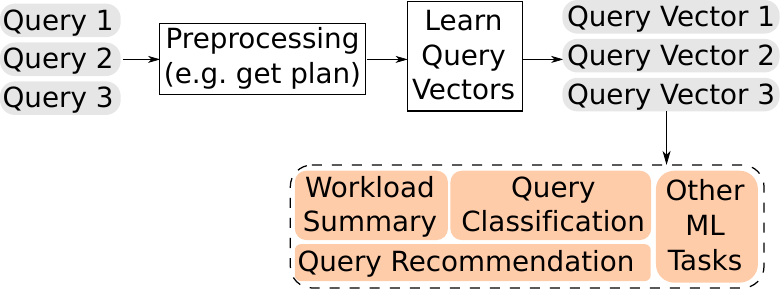}
\caption{A generic architecture for workload analytics tasks using embedded query vectors}\label{figure_architecture}
\end{figure}
To mine for patterns in large, unstructured data sources, data items must be represented in a standard form.  Representation learning \cite{bengio2013representation} aims to find semantically meaningful embeddings of semi-structured and unstructured data in a high-dimensional vector space to support further analysis and prediction tasks.  The area has seen explosive growth in recent years, especially in text analytics and natural language processing (NLP).  These learned embeddings are \emph{distributional} in that they produce dense vectors capable of capturing nuanced relationships --- the meaning of a document is distributed over the elements of a high-dimensional vector, as opposed to encoded in a sparse, discrete form such as bag-of-tokens.  In NLP applications, distributed representations have been shown to capture latent semantics of words, such that relationships between words correspond (remarkably) to arithmetic relationships between vectors.  For example, if $v(X)$ represents the vector embedding of the word $X$, then one can show that $v(\texttt{king}) - v(\texttt{man}) + v(\texttt{woman}) \approx v(\texttt{queen})$, demonstrating that the vector embedding has captured the relationships between gendered nouns.  Other examples of semantic relationships that the representation can capture include relating capital cities to countries, and relating the superlative and comparative forms of words \cite{mikolov2013distributed,mikolov2013efficient,le2014distributed}. 
Outside of NLP, representation learning has been shown to be useful for understanding nuanced features of code samples, including finding bugs or identifying the programmer's intended task \cite{mou2016convolutional}.


In this paper, we apply representation learning approaches to SQL workloads, with an aim of automating and generalizing database administration tasks.  
Figure \ref{figure_architecture} illustrates the general workflow for our approach.  For all applications, we consume a corpus of SQL queries as input, from which we learn a vector representation for each query in the corpus using one of the methods described in Section \ref{finding_vecs}.  We then use these representations as input to a machine learning algorithm to perform each specific task.

We consider two primary applications of this workflow: workload summarization for index recommendation \cite{chaudhuri2002compressing}, and identifying patterns of queries that produce runtime errors. Workload summarization involves representing each query with a set of specific features based on syntactic patterns.  These patterns are typically identified with heuristics and extracted with application-specific parsers: for example, for workload compression for index selection, Surajit et al. \cite{chaudhuri2002compressing} identify patterns like query type (SELECT, UPDATE, INSERT or DELETE), columns referenced, selectivity of predicates and so on.  Rather than applying domain knowledge to extract specific features, we instead learn a \emph{generic} vector representation, then cluster and sample these vectors to compress the workload. 

For query debugging, consider a DBA trying to understand the source of out-of-memory errors in a large cluster.  Hypothesizing that group by queries on high-cardinality, low-entropy attributes may trigger the memory bug, they tailor a regex to search for candidate queries among those that generated errors. But for large, heterogeneous workloads, there may be thousands of other hypotheses that they need to check in a similar way.  Using learned vector approaches, any syntactic patterns that correlate with errors can be found automatically. As far as we know, this task has not been considered in the literature, though we encounter this problem daily in production environments. We use real customer workloads from Snowflake Elastic Data Warehouse \cite{snowflake} for training and evaluating our model. We describe these applications in more detail in Section \ref{applications}. 

A key benefit of our approach in both applications is that a single pre-trained model can be used for a variety of applications. We show this effect empirically in Section \ref{sec:transferlearning}; we use a model trained on one workload to support an application on a completely unrelated workload. This transfer learning capability opens up new use cases, where a model can be trained on a large private workload but shared publicly, in the same way that Google shares pre-trained models of text embbeddings.

We make the following contributions:
\begin{itemize}  
\item We describe a general model for workload analysis tasks based on representation learning.
\item We adapt several NLP vector learning approaches to SQL workloads, considering the effect of pre-processing strategies.
\item We propose new algorithms based on this model for workload summarization and query error prediction. 
\item We evaluate these algorithms on real workloads from the Snowflake Elastic Data Warehouse \cite{snowflake} and TPC-H \cite{tpch}, showing that the generic approach can improve performance over existing methods.
\item We demonstrate that it is possible to pre-train models that generate query embeddings and use them for workload analytics on \textbf{unseen query workloads}.
\end{itemize}

This paper is structured as follows: We begin by discussing three methods for learning vector representations of queries, along with various pre-processing strategies (Section \ref{finding_vecs}).
Next, we present new algorithms for query recommendation, workload summarization, and two classification tasks that make use of our proposed representations (Section \ref{applications}).  We then evaluate our proposed algorithms against prior work based on specialized methods and heuristics (Section \ref{evaluation}).
We position our results against related work in Section \ref{related_work}. 
Finally, we present some ideas for future work in this space and some concluding remarks in Sections \ref{future_work} and \ref{conclusion} respectively.

\section{Learning Query Vectors}


\label{finding_vecs}
Representation learning \cite{bengio2013representation} aims to map some semi-struct\-ured input (e.g., text at various resolutions \cite{mikolov2013distributed,mikolov2013efficient,le2014distributed}, an image, a video \cite{video_representation}, a tree \cite{tree_lstm}, a graph\cite{anything2vec}, or an arbitrary byte sequence \cite{anything2vec, rudolph2016exponential}) to a dense, real-valued, high-dimensional vector such that relationships between the input items correspond to relationships between the vectors (e.g., similarity).

\begin{figure*}[ht]
  \centering
  \begin{subfigure}{0.31\textwidth}
  \includegraphics[width=1\columnwidth]{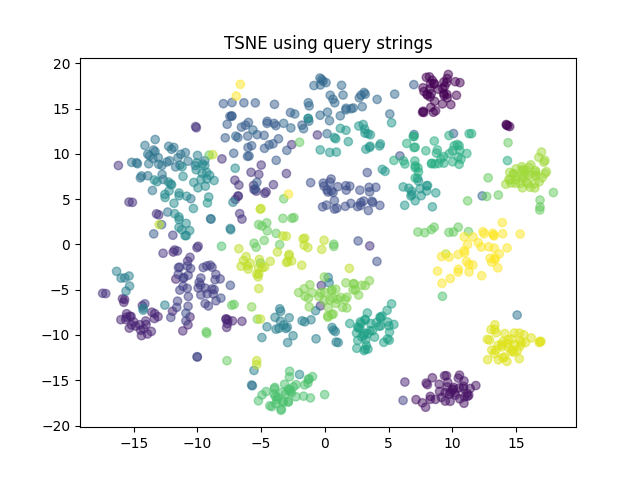}
  \caption{Embedded queries}
  \label{tsne-tpch-queries}
  \end{subfigure}
  \begin{subfigure}{0.31\textwidth}
    \includegraphics[width=1\textwidth]{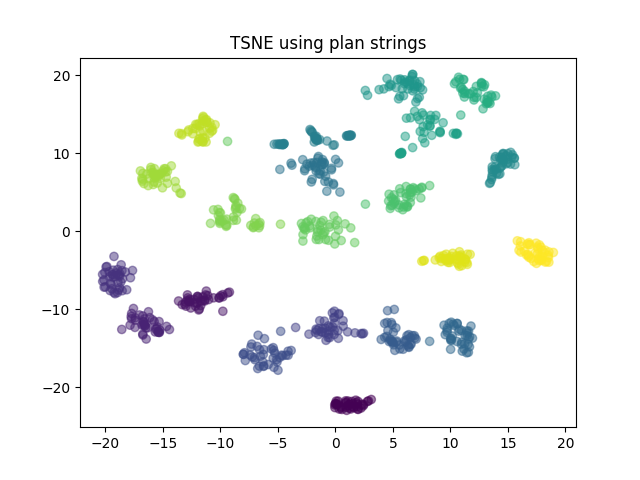}
    \caption{Embedded query plans}
    \label{tsne-tpch-plans}
  \end{subfigure}  
  \begin{subfigure}{0.31\textwidth}
  \includegraphics[width=1\columnwidth]{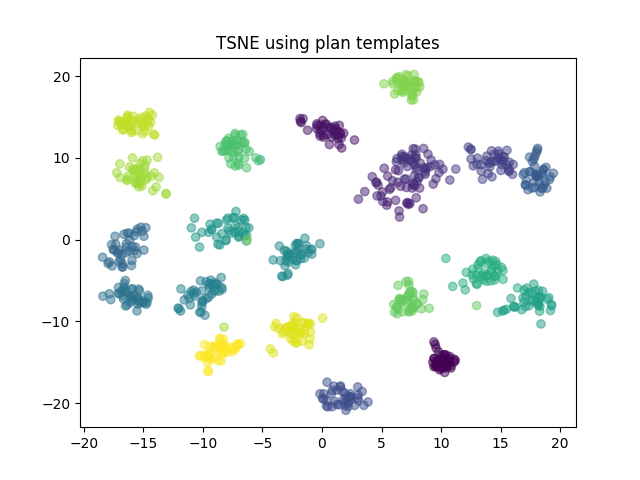}
  \caption{Embedded query plan templates}
  \label{tsne-tpch-plan_templates}
  \end{subfigure}
  \caption{Different embedding strategies exhibit different clustering behavior.  We considered three embeddings for the TPC-H workload and then cluster the results in 2D using the T-SNE algorithm.   Each point on the graph is a single query, and the color represents the original TPC-H query template from which the query was generated. (a) Embedding the raw SQL produces qualitatively less cohesive clusters.  (b, c) Embedding plans and plan templates produces more distinct clusters.  But in all three cases, the learned embeddings are able to recover the known structure in the input. }
  \label{TSNE-tpch}
\end{figure*}
For example, NLP applications seek vector representations of textual elements (words, paragraphs, documents) such that semantic relationships between the text elements correspond to arithmetic relationships between vectors.  

\paragraph*{Background} As a strawman solution to the representation problem for text, consider a one-hot encoding of the word \texttt{car}: A vector the size of the dictionary, with all positions zero except for the single position corresponding to the word \texttt{car}.  In this representation, the words \texttt{car} and \texttt{truck} are as dissimilar (i.e., orthogonal) as are the words \texttt{car} and \texttt{dog}.

A better representation is suggested by the \emph{distributional hypothesis}~\cite{sahlgren:08}: that the meaning of a word is carried by the contexts in which it appears.  To take advantage of this hypothesis, one can set up a prediction task: given a context, predict the word (or alternatively, given a word, predict its context.)  The weights learned by training a neural network to make accurate predictions become an alternative representation of each word: a vector that captures all of the contexts in which the word appears.  This approach has led to striking results in NLP, where semantic relationships between words can be shown to approximately correspond to arithmetic relationships between vectors \cite{pennington2014glove}.

In this section, we discuss alternatives for adapting representation learning for SQL queries.  We first discuss pre-processing strategies, including removing literals and the tradeoffs between using raw SQL or optimized plans as inputs.  Next we describe a direct application of Doc2Vec by predicting each token in the query from its context, i.e. ``continuous bag-of-words" \cite{mikolov2013efficient,mikolov2013distributed}). Finally we use Long Short Term Memory networks (LSTMs) to implement an autoencoder; this approach allows us to avoid explicitly defining the size of the context to be used and thereby better adapt to heterogeneous queries.

\paragraph*{Preprocessing: Query Plan Templates}

The optimized plan rather than the SQL text can be used as input to the network, but this strategy has strengths and weaknesses.  There are variations in how a query may be expressed in SQL (e.g., expressing a join using \texttt{JOIN ... ON ...} vs. the \texttt{WHERE} clause).  For some workload analysis tasks, these variations may become confounding factors, since two equivalent queries can produce very distinct representations. For example, deriving a set of indexes to improve performance for a given workload has little to do with how the SQL was authored, and everything to do with the plan selected by the optimizer. However, for other tasks, the ability to capture patterns in how users \emph{express} queries as opposed to how the database \emph{evaluates} the queries can be important.  For example, query recommendation tasks that aim to improve user productivity \cite{querie} and should therefore be responsive to stylistic differences in the way various users express queries.  

We remove literal values in the plan before training the network, primarily because the network will not be able to learn numeric relationships.  We do not remove attribute and table identifiers: the usage patterns of specific tables and attributes in a workload may be informative.
In some multi-tenant situations, table names and attribute names can also help identify patterns even across different customer schemas. For example, in the SQLShare workload \cite{shrjainSQLShare} we found that the table name \texttt{phel\_clc\_blastx\_uniprot \_sprot\_sqlready\_1.tab} was considered most similar to the table name \texttt{phel\_deseq2\_sig\_results \_c.sig} in one data sharing workload.  These tables correspond to different users and are not otherwise obviously related, but they were routinely queried the same way in the workload.  Investigating, we found that these tables were uploaded by different users in the same community who were performing similar experiments.

After removing literals and constants, we obtain an optimized query plan from the database (or logs), which also contains estimated result sizes and estimated runtimes for each operator.  The plan is represented as an XML document.  We remove angle brackets and other syntactic features from the XML before learning a representation, but we found that this transformation did not significantly affect the results.


We now explore different strategies for learning representations for queries: A learned representation approach using context windows of a fixed size (i.e., Doc2Vec), and an approach based on LSTMs which automatically learns an appropriate context window (more precisely: the relative influence of tokens in the surrounding context is itself learned.) All of these approaches can work on either the raw query string or the processed query plan template. 

\subsection{Method 1: Doc2Vec}

\begin{figure}[h]
  \centering
   \includegraphics[width=\columnwidth]{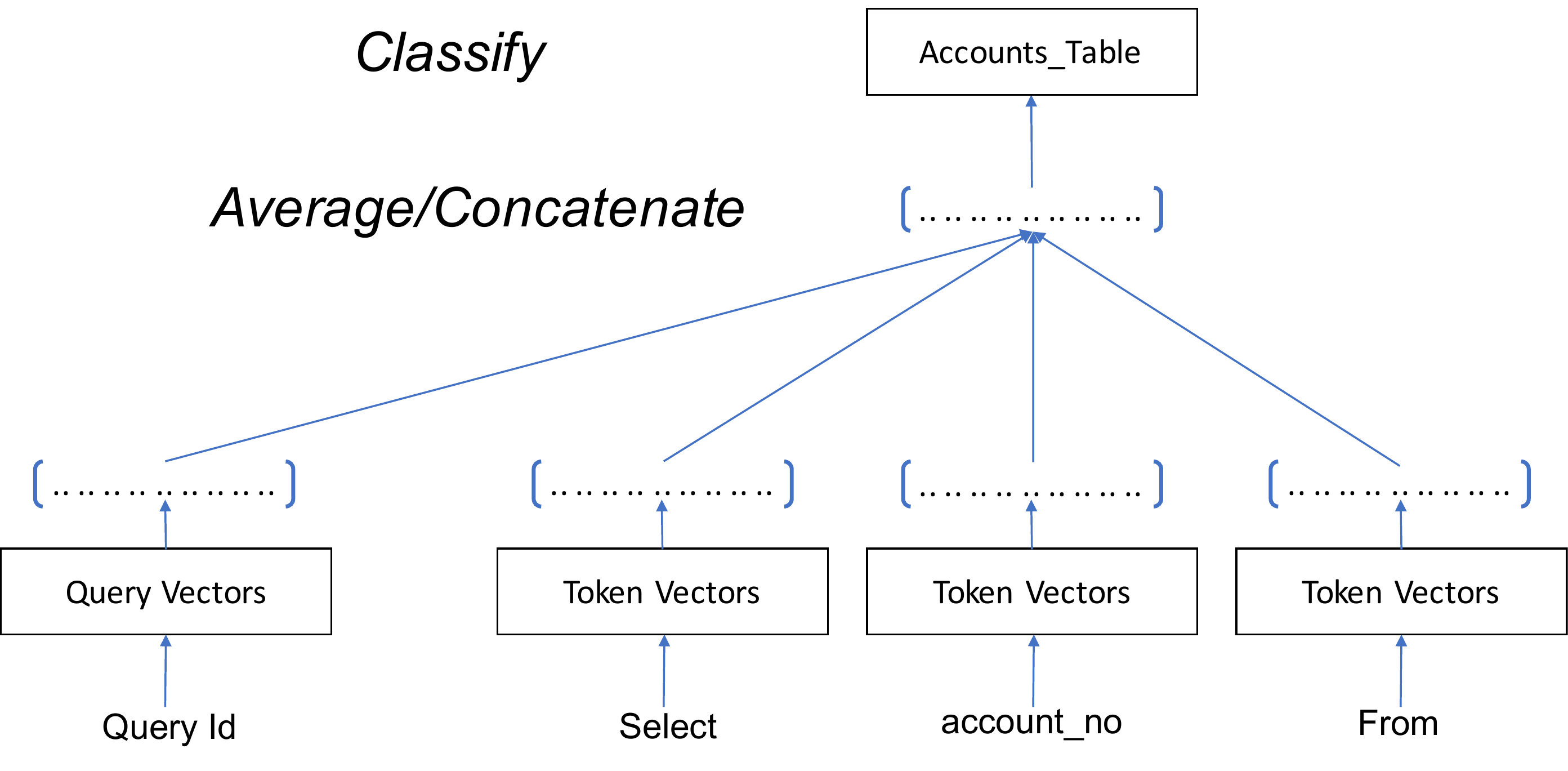}
    \caption{Architecture for learning query vectors using the Doc2Vec algorithm (we use a similar figure as \protect\cite{le2014distributed} for the sake of clarity). The algorithm uses the context tokens and the query id to predict the next token in the query and uses back propagation to update these token vectors and query vectors.}
    \label{doc2vec_image}
  \end{figure}

To improve on previous token-frequency based representations, Mikolov et al. \cite{mikolov2013distributed,mikolov2013efficient,le2014distributed} learn a vector representation for words in a document corpus by predicting the next word in context, then derive a vector representation for larger semantic units (sentences, paragraphs, documents) by adding a vector representing the paragraph to each context as an additional ``word" to provide memory across context windows (see Figure \ref{doc2vec_image}). The learned vector for this virtual word is used as a representation for the entire paragraph. Vectors learned in this manner have been shown to work remarkably well for sentiment classification and clustering tasks \cite{convnetNNsentenceClassification, lecunmnist}.

This approach can be applied directly for learning representations of queries: We can use fixed-size context windows to learn a representation for each token in the query, and include a query identifier to learn a representation of entire query.  This approach implies a hypothesis that the semantics of a query is an aggregation of the semantics of a series of local contexts; evaluating this hypothesis for workload analytics tasks is one of the goals of this paper.

\begin{algorithm}
\caption{Query2Vec-Doc2vec}
\label{q2vd2v_algo}
	\begin{algorithmic}[1]  
    	\STATE \textbf{Input}: A query workload $W$ and a window size $k$
        \STATE Initialize a random vector for each unique token in the corpus.
        \STATE Initialize a random vector for each query in workload.
        \FORALL{$q \in W$}
        	\STATE $q_p \leftarrow \textsf{preprocess}(q)$ 
            \STATE $q_v \leftarrow$ replace each token in $q_p$ with token vector
            \STATE $v_q \leftarrow$ initialized vector corresponding to $q$ 
            \FOR{each context window $(v_0, v_1, ..., v_k$) in $q_v$}
            \STATE predict the vector $v_k$ using  $(v_q + v_0 + v_1 + ... v_{k-1}) \div k$
            \ENDFOR
            \STATE compute errors and update $v_q$ via backpropagation
        \ENDFOR
        \STATE \textbf{Output}: Learned query vectors $v_q$
\end{algorithmic}
\end{algorithm}

Although the local contexts are defined assuming a simple sequence of tokens (the previous $k$ tokens and the subsequent $k$ tokens), this approach is not necessarily inappropriate even for tree-structured inputs like queries: local sequential contexts are well-defined in any reasonable linearization of the tree.  In this paper, we linearize the query plan and query plan templates using an in-order traversal. We describe this process in Algorithm \ref{q2vd2v_algo}.

To qualitatively evaluate whether these learned vector representation for queries can recover semantics, we can generate embeddings for queries with known patterns and see if we can recover those patterns in the derived vector space.  In Figure \ref{TSNE-tpch}, we show clusters derived from embedding (a) the raw SQL, (b) the query plan, and (c) the templatized query plan.  Using the 21 TPC-H queries as patterns, we generated 200 random instances of each TPC-H query pattern.  Next we learned vector embeddings model on raw query strings (\ref{tsne-tpch-queries}) and generated 300 dimensional vectors for each query. We then used t-SNE \cite{maaten2008visualizing} to reduce these embeddings to two dimensions in order to visualize them.  We repeated the same experiments on linearized query plans (\ref{tsne-tpch-plans}) and linearized query plan templates (\ref{tsne-tpch-plan_templates}). The color of each point represents the ground truth of the original TPCH query type, unknown to the model. All algorithms are able to recover the original structure in this contrived workload, but the clusters resulting from the raw SQL are less cohesive, suggesting that the use of optimized plans may be more effective.

\subsection{Method 2: LSTM Autoencoder}

\begin{figure}
  \centering
    \includegraphics[width=0.5\textwidth]{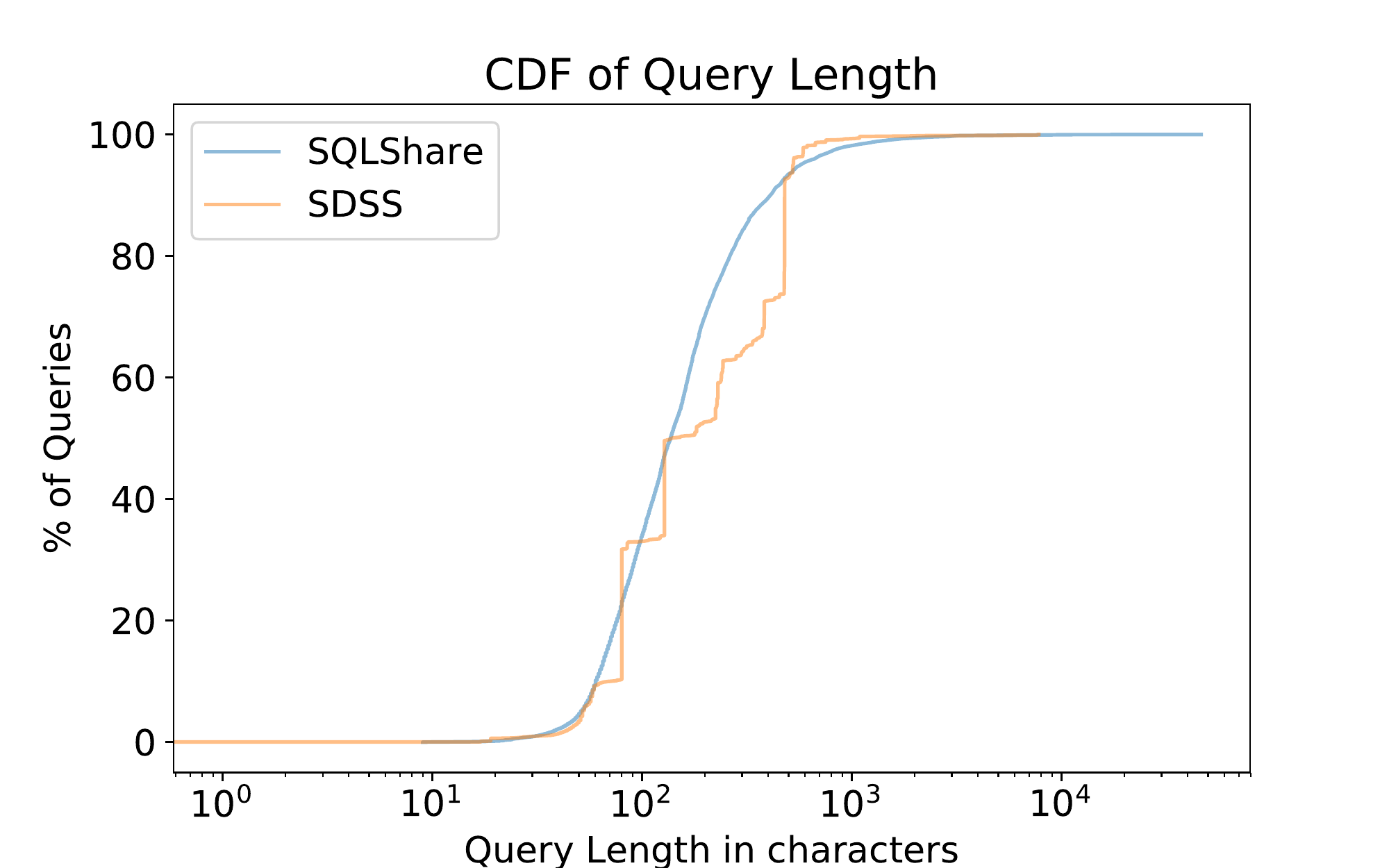}
    \caption{Distribution of query length across two real SQL worklaods, the figure shows that query length varies widely in ad-hoc workloads.}
    \label{sqlshare_sdss_length}
  \end{figure}

The paragraph vector approach in the previous section is viable, but it requires a hyper-parameter for the context size. There is no obvious way to determine a fixed context size for queries, for two reasons: First, there may be semantic relationships between distant tokens in the query.  As an illustrative example, attribute references in the \texttt{SELECT} clause correspond to attribute references in  the \texttt{GROUP BY} clause, but there may be arbitrary levels of nesting in between.  Second, the length of queries vary widely in ad hoc workloads \cite{shrjainSQLShare,sqlshare_data}.
Figure \ref{sqlshare_sdss_length} illustrates the query length distribution for two real SQL workloads \cite{sqlshare_data}. 

To avoid setting a context size, we can use Long Short-Term Memory (LSTM) networks, which are modified Recurrent Neural Networks (RNN) that can automatically learn how much context to remember and how much of it to forget, thereby removing the dependence on a fixed context size (and avoid the \emph{exploding} or \emph{vanishing} gradient problem~\cite{Hochreiter01gradientflow}). LSTMs have successfully been used in sentence classification, semantic similarity between sentences and sentiment analysis \cite{tang2015document}. 

\paragraph*{Background: Long Short Term Memory (LSTM) networks}
Zaremba and Sutskever define an LSTM unit at time $t$ (time is equivalent to position of tokens in the processed query) to contain a memory cell $c_t$, a hidden state $h_t$, an input gate $i_t$, a forget gate $f_t$ and an output gate $o_t$ \cite{zaremba2014recurrent}.  The \emph{forget gate} scales the output from the previous cell output \emph{$h_{t-1}$} which is passed to the memory cell at time \emph{t}. 
The transition equations governing these values are as follows:

\begin{equation} \label{seq_lstm}
\begin{split}
i_{t}	&=   \sigma \left( W^{(i)} x_{t}  +  U^{(i)} h_{t-1} +  b^{(i)} \right) \\
f_{t}	&=   \sigma \left( W^{(f)} x_{t}  +  U^{(f)} h_{t-1} +  b^{(f)} \right)\\
o_{t}	&=   \sigma \left( W^{(o)} x_{t}  +  U^{(o)} h_{t-1} +  b^{(o)} \right)\\
u_{t}	&=   \tanh \left( W^{(u)} x_{t}  +  U^{(u)} h_{t-1} +  b^{(u)} \right)\\
c_t 	&=	i_t \odot u_t + f_t \odot c_{t-1}\\
h_t		&= 	o_t \odot \tanh\left(c_t\right)\\
\end{split}
\end{equation}
\noindent
where \emph{x} is the input at current time step, $\sigma$ denotes logistic sigmoid function and $\odot$ denotes the \emph{Hadamard} product. The matrices $W$ and $U$ represent weights and vector $b$ represents the bias. The superscript $i$, $f$, $o$ and $u$ in the weight and bias matrices represent the corresponding weight matrices and biases for input gate, forget gate, output gate and cell state respectively. These are the parameters that are learned by the network using backpropagation \cite{backprop-hinton}.
The state passed from one cell to the next are the cell state $c_i$ and the hidden state $h_i$. The hidden state $h_i$ represents the an encoding of the input seen until the $i^{th}$ cell.

    

\begin{figure*}
  \centering
  \begin{subfigure}[b]{0.7\textwidth}
  \includegraphics[width=\textwidth]{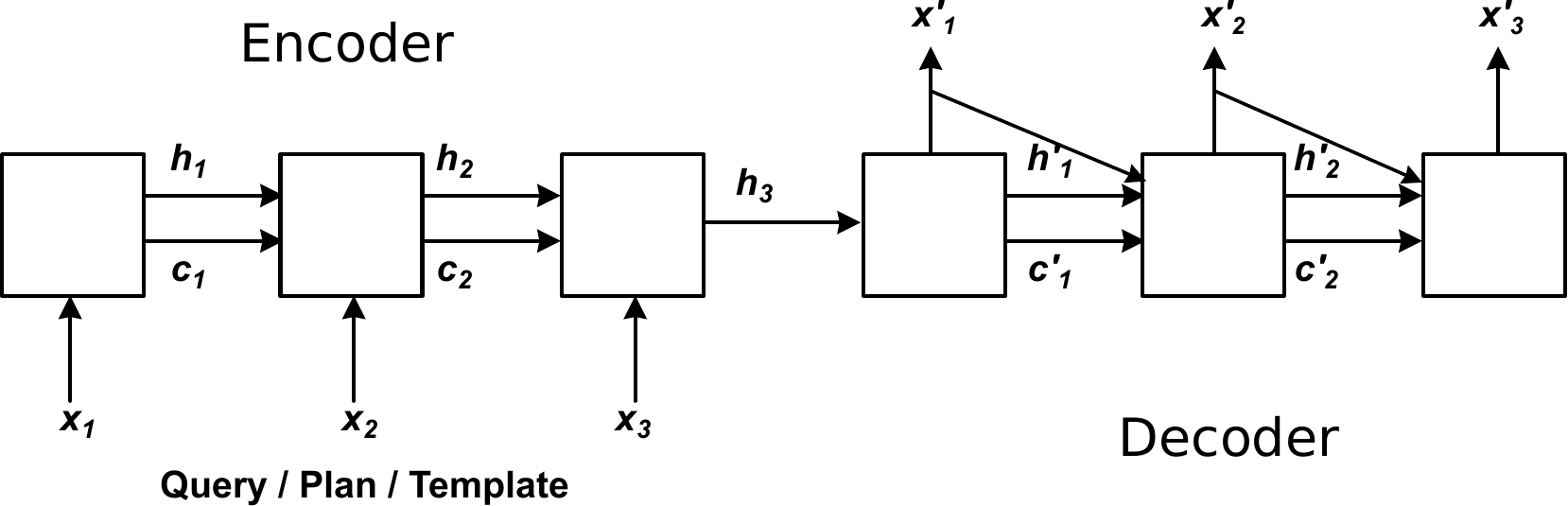}
 
  \end{subfigure}
  \caption{Long Short Term Memory (LSTM) Autoencoder based Query2Vec architecture. The encoder LSTM network models the query as a sequence of words (figure shows an example query with 3 tokens) and uses the hidden state of the final encoder LSTM cell as an input to decoder LSTM which reproduces the input query. Once the network is trained, query vectors can be inferred by passing the query to the encoder network, running a forward pass on the LSTM components and outputting the hidden state of the final encoder LSTM cell as the inferred vector.}
  \label{seq_lstm_fig}
\end{figure*}  

\paragraph*{LSTM based autoencoder}
We use a standard LSTM encoder decoder network \cite{auto_encoders,lstm_autoencoders} with architecture as illustrated in 
Figure \ref{seq_lstm_fig}. A typical autoencoder trains the network to reproduce the input. The encoder LSTM part of the network encodes the query (or query plan), and this encoding is subsequently used by the later half of the network (decoder) which tries to reproduce the input query (or query plan). More concretely, we train the network by using the hidden state of the final LSTM cell (denoted by $h_{n}$ in the figure) to reproduce the input query (or query plan). The decoder network outputs one token at a time, and also uses that as an input to the subsequent LSTM cell to maintain context.

Our LSTM-based approach is given in Algorithm \ref{q2vlstm_algo}. An LSTM autoencoder is trained by sequentially feeding words from the pre-processed queries to the network one word at a time, and then reproduce the input. The LSTM network not only learns the encoding for the samples, but also the relevant context window associated with the samples as the forget gate scales the output through each step. The final hidden state ($h_i$) on the encoder network gives us an encoding for the samples.

Once this network has been trained, an embedded representation for a query can be computed by passing the query to the encoder network, completing a forward pass, and using the hidden state of the final encoder LSTM cell as the learned vector representation. Lines 9 through 13 in Algorithm \ref{q2vlstm_algo} describe this process procedurally.
 
\begin{algorithm}
\caption{Query2vec-LSTM}
\label{q2vlstm_algo}
	\begin{algorithmic}[1] 
      \STATE \textbf{Input}: A query workload $W$.
        \STATE Assign random weights to each token in query corpus
        \FOR{each query $q \in W$}
            \STATE $q_p \leftarrow$ preprocess($q$)
            \STATE pass the token vectors through the encoder LSTM network
            \STATE generate the tokens back using the hidden state of the final encoder LSTM cell as input to the decoder LSTM network
            \STATE compute error gradient and update network weights via backpropagation
        \ENDFOR
        \STATE Initialize learned vectors $V$ to an empty list
         \FOR{each query $q \in W$}
        	\STATE pass $q$ through the learned encoder network
            \STATE Append the hidden state of the final encoder LSTM cell to $V$
        \ENDFOR
        \STATE \textbf{Output}: learned vectors $V$
\end{algorithmic}
\end{algorithm}

\section{Applications}

\label{applications}

In this section, we use the algorithms from Section \ref{finding_vecs} as the basis for new algorithms for workload analysis tasks. We will evaluate these applications in Section \ref{evaluation}.  We consider two applications in detail: workload summarization for index selection \cite{chaudhuri2002compressing} and resource error prediction. Then, in Section \ref{future_work}, we consider some novel applications enabled by these Query2Vec representations.

In each case, a key motivation is to provide services for a multi-tenant data sharing system, where many users are working with many schemas. In these situations, it is difficult to exploit heuristics to identify workload features since the schemas, query ``style,'' and application goals can vary widely.  Learned vector representations can potentially offer a generic approach to workload analytics that adapts automatically to patterns in the workload, even across schemas, users, and applications.  
Our underlying hypothesis is that a common representation of queries provides a basis for algorithms for a variety of workload analysis tasks,and that these new algorithms can compete favorably with more specialized algorithms tailored for each specific task.
We use the architecture proposed in Figure \ref{figure_architecture} for the applications that follow in the later parts of this section.

In this paper, we generally assume that we will train the model on a corpus that is specific to the application or system we are trying to support.  However, we consider it likely that one can pre-train models on a large shared public query corpus (e.g., SDSS \cite{singh2007skyserver} or SQLShare \cite{shrjainSQLShare}), and then re-use the pre-trained vectors for specific analysis tasks, in the same way that NLP applications and image analysis applications frequently re-use models and learned vectors pre-trained on large text and image corpora. We demonstrate some preliminary results of this approach in Section \ref{sec:transferlearning} and leave an analysis of the tradeoffs of this transfer learning approach to future work.

\subsection{Workload Summarization}
\label{sec_ws}
The goal of workload summarization \cite{chaudhuri2002compressing, kolaczkowski2008compressing} is to find a representative sample of the whole workload as input to further database administration and tuning tasks, including index selection, approximate query processing, histogram tuning, and statistics selection \cite{chaudhuri2002compressing}.  We focus here on index recommendation as a representative application of workload summarization.  The goal is to find a set of indexes that will improve performance for a given workload \cite{chaudhuri1997efficient}.  Most commercial database systems include index recommendation as a core feature; Microsoft SQL Server has included the feature as part of its Tuning Advisor since SQL Server 2000 \cite{agrawal:05}.  We will not further discuss the details of the index recommendation problem; following previous work on workload summarization, we make the assumption that it is a black box algorithm that accepts a workload as input and produces a set of indexes as an output.  

The workload summarization task (with respect to index recommendation) is as follows:  Given a query workload $\mathcal{Q}$, find a representative subset of queries $\mathcal{Q}_{sub}$ such that the set of indexes recommended based on $\mathcal{Q}_{sub}$ approximate the set of indexes of the overall workload $\mathcal{Q}$.  Rather than attempt to measure the similarity of two sets of indexes (which can vary in syntactic details that may or may not significantly affect performance), we will follow the literature and measure the performance of the overall workload as the indicator of the quality of the indexes.  Workload summarization is desirable not only to improve recommendation runtime (a roughly $O(N^2)$ task), but also potentially to optimize query runtime performance during ordinary database operations \cite{chaudhuri2002compressing}.  


Previous techniques for workload summarization are primarily variants of workload summarization algorithms described by Chaudhury et al.  \cite{chaudhuri2002compressing}. 
They describe three main approaches to the problem: 1) a stratified random sampling approach, 2) a variant of K-Mediod algorithm which produces $K$ clusters and automatically selects an appropriate $K$ and 3) a greedy approach which prunes the workload of queries which do not satisfy the required constraints. Case 2) is highly dependent on the distance function between queries; the authors emphasize that a custom distance function should be developed for specific workloads.  As we have argued, this approach is increasingly untenable for multi-tenant situations with highly dynamic schemas and workloads.  Of these, the K-Mediod approach, equipped with a custom distance function that looks for join and group by patterns, performs the best in terms of quality of compression and time taken for compression. 


Our observation is that the custom distance function is unnecessary:  we can embed each query into a high-dimensional vector space in a manner that preserves its semantics, such that simple cosine distance between the vectors performs as well as application-specific distance functions over applications-specific feature vectors.  

Since we no longer need the custom distance function, we can also use the much faster k-means algorithm rather than the more flexible (but much slower) K-Mediods algorithm originally proposed: K-Mediods selects an element in the dataset as the centroid, meaning that distances are always computed between two elements as opposed to between an element and the mean of many elements.  The upshot is that K-Mediods supports any arbitrary distance function, a feature on which the original method depends crucially since these functions are black boxes.  We do still need to select a member of the cluster to return at the last step; we cannot just return the centroid vector, since an arbitrary vector cannot be inverted back into an actual SQL query.  Therefore we perform an extra linear scan to find the nearest neighbor to the computed centroid.

We illustrate our algorithm in figure \ref{ws} and provide a procedural description in Algorithm \ref{sql_summary_algo}. We use  K-means to find $K$ query clusters and then pick the closest query to the centroid in each cluster as a representative query for that cluster in the summary. To determine the optimal $K$ we use the Elbow method \cite{kodinariya2013review} which runs the K-means algorithm in a loop with increasing $K$ till the rate of change of the sum of squared distances from centroids `elbows' or plateaus out. We use this method due to its ease of implementation, but our algorithm is agnostic to the choice of the method to determine $K$. 

\begin{algorithm}
\caption{Q2VSummary: Summarize a SQL Workload}
\label{sql_summary_algo}
\begin{algorithmic}[1]
    \STATE Train Query2Vec model. (optional one time step, can be offline)
    \STATE Use Query2Vec to convert each query $q$ to a vector $v_q$. 
    \STATE Determine $k$ using the elbow method \cite{kodinariya2013review}.
    \STATE Run $k$-means on the set of query vectors $v_0, v_1, ..., v_n$.
    \STATE For each cluster $k$, find the nearest query to the centroid via linear scan using cosine distance.
    \STATE Return these $k$ queries  as the summarized workload $\mathcal{Q}_{sub}$.
\end{algorithmic}
\end{algorithm}

\subsection{Classification by Error}
\label{errorforensics}

An individual query is unlikely to cause a mature relational database system to generate an error.  However, modern MPP databases and big data systems are significantly more complex than their predecessors and operate at significantly larger scale.  In the real workload we study (from Snowflake Elastic Data Warehouse \cite{snowflake}), from a 10-day window from one deployment consisted of about $70$ million \emph{select} queries, about 100,000 queries resulted in some kind of error ($0.19\%$). Moreover, systems have become increasingly centralized (e.g., as data lakes) and multi-tenant (e.g., cloud-hosted services.)  In this heterogeneous regime, errors are more common, and identifying the syntactic patterns that tend to produce errors can not be performed manually.

\begin{figure*}[ht]
  \begin{subfigure}{\textwidth}
  \centering
  \includegraphics[width=0.85\columnwidth]{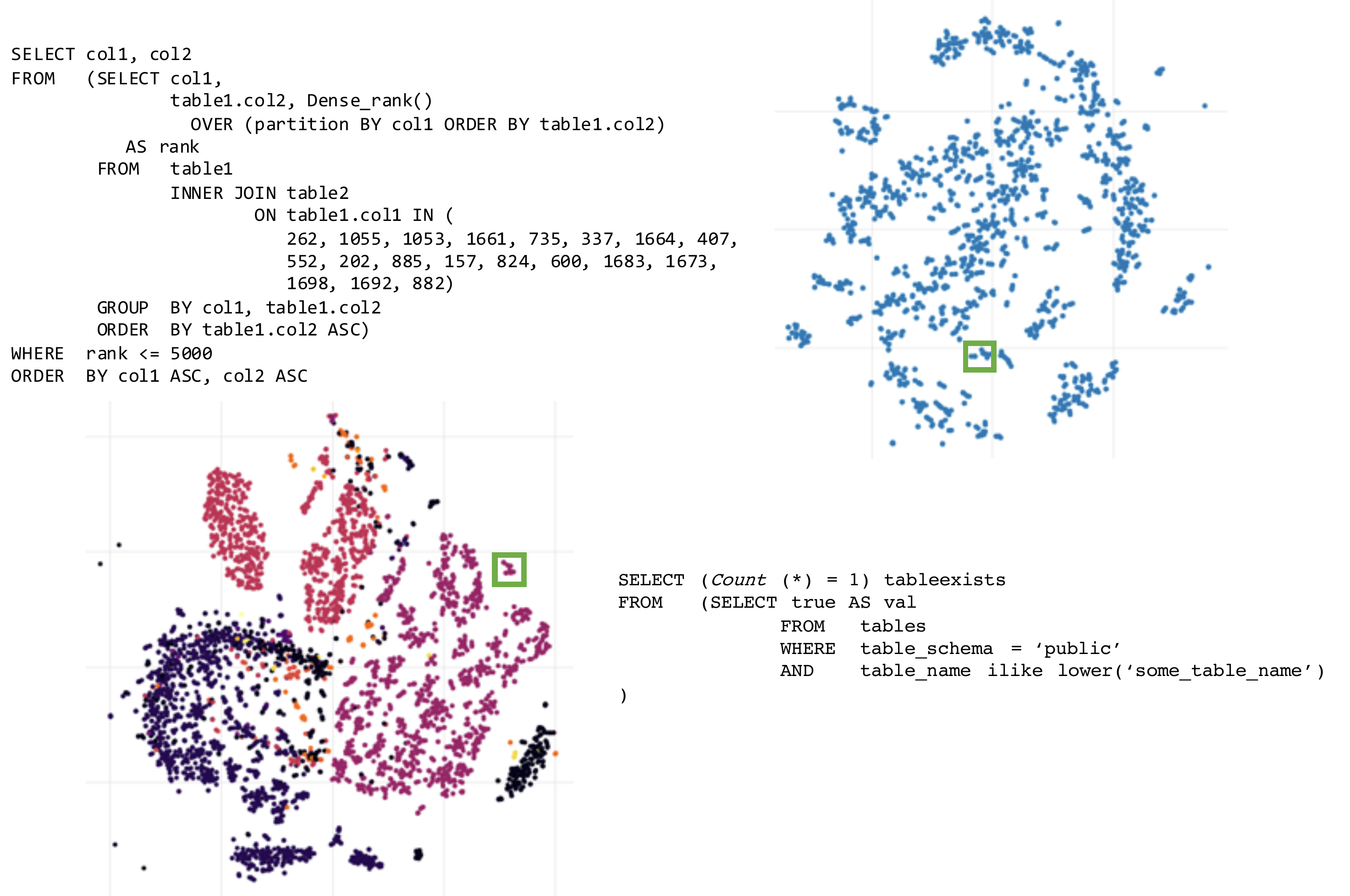}
  \end{subfigure}
  \caption{A clustering of error-generating SQL queries from a large-scale cloud-hosted multi-tenant database system. Color represents the type of error. The syntax patterns in the workload are complex as one would expect, but there are obvious clusters some of which are strongly associated with specific error types. The text on the left and the right of the plots correspond to the queries annotated in the plots respectively.}
  \label{fig:errormap}
\end{figure*}

In figure \ref{fig:errormap}, we show a clustering of error-generating SQL queries from a large-scale cloud-hosted multi-tenant database system.  Color represents the type of error; there are over 20 different types of errors ranging from out-of-memory errors to hardware failures to query execution bugs. The syntax patterns in the workload are complex (as one would expect), but there are obvious clusters, some of which are strongly associated with specific error types.  For example, the cluster at the upper right corresponds to errors raised when a timeout was forced on dictionary queries.

Using an interactive visualization based on these clusterings, the analyst can inspect syntactic clusters of queries to investigate problems rather than inspecting individual queries, for two benefits: First, the analyst can prioritize large clusters that indicate a common problem.  Second, the analyst can quickly identify a number of related examples in order to confirm a diagnosis.  For example, when we first showed this visualization to our colleagues, they were able to diagnose the problem associated with one of the clusters immediately.

\section{Evaluation}
\label{evaluation}

We evaluate the methods for learning vector representations described in Section \ref{finding_vecs} against the applications described in Section \ref{applications}. 

\subsection{Experimental Datasets}
\paragraph*{Datasets for training query2vec}
We evaluate our summarization approach against TPC-H in order to facilitate comparisons with experiments in the literature, and to facilitate reproducibility.  We use scale factor 1 (dataset size 1GB) in order to compare with prior work.   We generated queries with varying literals for 21 different TPC-H query templates.  We ignored query 15 since it includes a \texttt{CREATE VIEW} statement. 
 
We evaluate error forensics using a real workload from Snowflake \cite{snowflake}, a commercial large-scale, cloud-hosted, multi-tenant database service.  This service attracts millions of queries per day across many customers, motivating a comprehensive approach to workload analytics.  For training the query2vec algorithms, we used a random sample with $500000$ \emph{select} queries over a 10-day window of all customer queries. The following query over the jobs table in the commercial database service was used:

\begin{verbatim}
select * from (
    select * from (
        select query, id 
        from jobs -- logs table
        where created_on > '1/09/2018'
          and statement_properties = 'select' 
    ) tablesample (16) -- random sample
) limit 500000;
\end{verbatim}

The statistics for these workloads appear in Table \ref{dataset_details}.
\begin{table}
\centering
\begin{tabular}{|l|L{2cm}|L{1.5cm}|}
\hline
Workload& Total Queries & Distinct  \\ \hline
Snowflake & 500000 & 175958 \\ \hline
TPC-H & 4200 & 2180\\ \hline
\end{tabular}
\caption{Workloads used to train query2vec}
\label{dataset_details}
\end{table}

\paragraph*{Datasets for experimental evaluation}
For workload summarization experiment, we evaluate the query2vec model (trained on the TPC-H dataset in Table \ref{dataset_details}) by generating summary for another subset of $840$ TPC-H queries.
For error forensics, we use 2 datasets. First dataset contains an even mix of queries that failed and queries with no errors (Snowflake-MultiError). The Second dataset contains queries that failed due to out-of-memory errors and queries with no errors.
Details for these evaluation workloads appear in Table \ref{eval_dataset_details}.
\begin{table}
\centering
\begin{tabular}{|l|L{2cm}|L{1.5cm}|}
\hline
Workload& Total Queries & Distinct \\ \hline
Snowflake-MultiError & 100000 & 17311 \\ \hline
Snowflake-OOM & 4491 & 2501 \\ \hline
TPC-H & 840 & 624\\ \hline
\end{tabular}
\caption{Datasets used for evaluation}
\label{eval_dataset_details}
\end{table}



\subsection{Workload Summarization and Indexing}
\label{sec:summary-results}
In this section we measure the performance of our workload summarization algorithm. Following the evaluation strategy of Chaudhuri et al.\cite{chaudhuri2002compressing}, we first run the index selection tool on the entire workload $\mathcal{Q}$, create the recommended indexes, and measure the runtime $t_{orig}$ for the original workload.    Then, we run our workload summarization algorithm to produce a reduced set of queries $\mathcal{Q}_{sub}$, re-run the index selection tool, create the recommended indexes, and again measure the runtime $t_{sub}$ of the entire original workload.  

We have performed experiments to directly compare with prior work and found significant benefits, but since the original paper reports on results using a much earlier version of SQL Server, the results were difficult to interpret.  Since the earlier compression techniques have since been incorporated directly into SQL Server \cite{microsoft-technet}, we instead use our method as a pre-processing step to see if it adds any additional benefit beyond the state-of-the-art built-in compression approach.  The documentation for SQL Server states that workload compression is applied; we did not find a way to turn this feature off.

We further apply a time budget to show how giving the index recommendation tool more time produces higher quality recommendations and improved performance on the workload.

For this experiment, we use SQL Server 2016 which comes with the built-in Database Engine Tuning Advisor. We set up the SQL Server on a $m4.large$ AWS EC2 instance. We set up TPC-H with scale factor 1 on this server. The experimental workflow involved submitting the workload (or summary) to Database Engine Tuning Advisor (with varying time budget), getting recommendations for the indexes, creating these indexes, running the complete TPC-H workload (Table \ref{eval_dataset_details}) and measuring the total time taken, clearing the cache and finally dropping the indexes (for the next run of the experiment).

\begin{figure}
\centering
\includegraphics[keepaspectratio=true, width=\columnwidth]{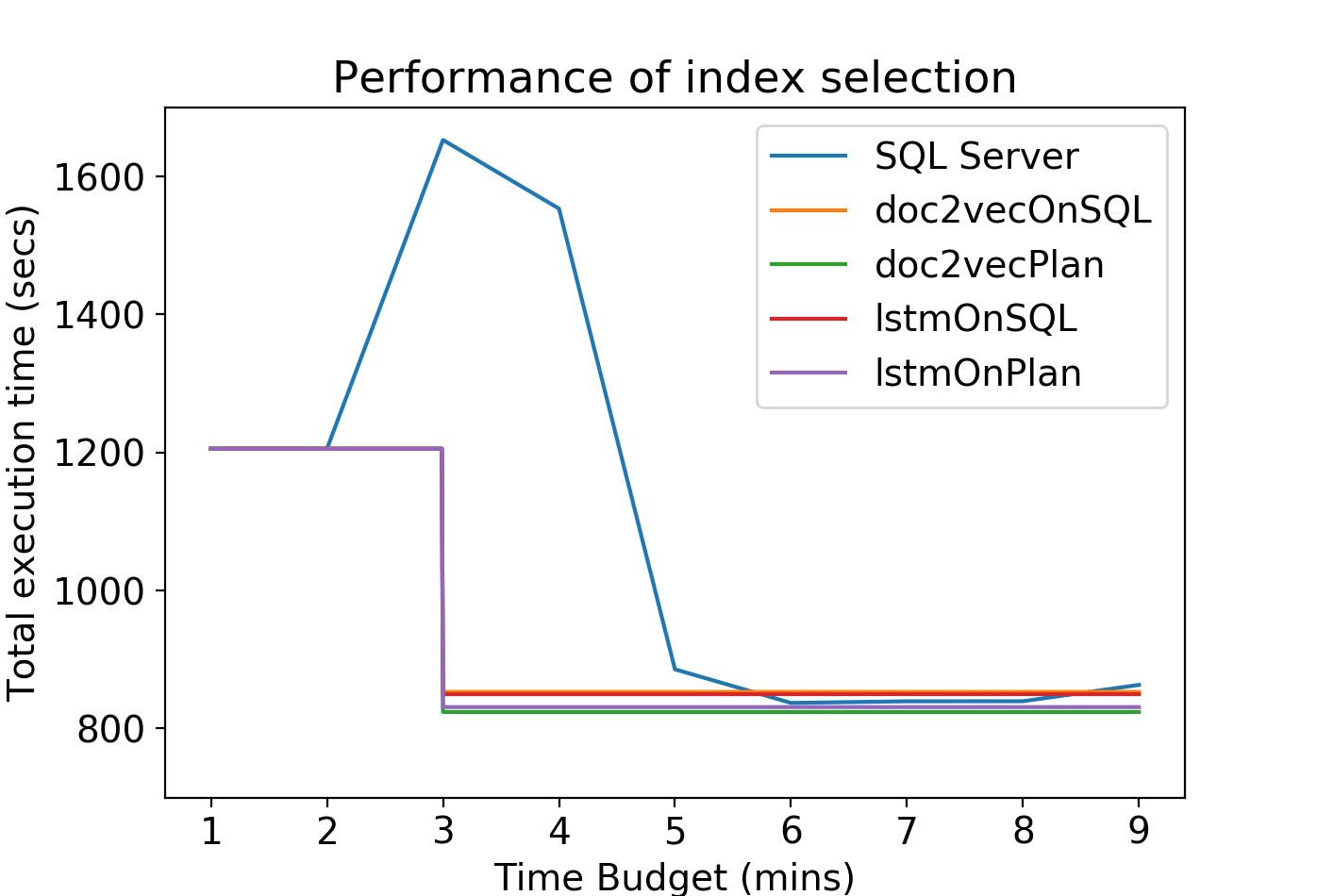}
\caption{Workload runtime using indexes recommended under various time budgets.  Using a query2vec compression scheme, the indexes recommended resulted in improved runtimes under all time budgets over 3 minutes.  Under 3 minutes, our method does not finish, and we default to producing no index recommendations. Surprisingly, the recommendations for the uncompressed workload results in \emph{worse} runtimes under tight time budgets before converging at higher time budgets.  These results show that syntax-based workload compression as a pre-processing can improve the performance of index recommenders.}
\label{fig:summarization}
\end{figure}

Figure \ref{fig:summarization} shows the results of the experiment. The x-axis is the time budget provided to the index recommender (a parameter supported by SQL Server).  The y-axis is the runtime for the entire workload, after building the recommended indexes.  The times for the learned vector embedding methods are bi-modal: For time budgets less than  3 minutes, our methods do not finish (and SQL server produces no recommendations as well), and we default to recommending zero indexes.  These numbers do not include query2vec training time: in practice, the model would only be trained once and used for a variety of applications.  This time only includes the time to process the queries to produce a learned vector representation.  This step is still expensive due to the multi-layer neural network used (e.g., a chain of vector-matrix multiply operations must be performed.)  We have not attempted to optimize this step, though it is trivial to do so: the workload can be processed in parallel, and a number of high-performance engines for neural networks are available.  

\begin{figure*}[ht]
  \centering
  \begin{subfigure}{0.45\textwidth}
  \includegraphics[width=1\columnwidth]{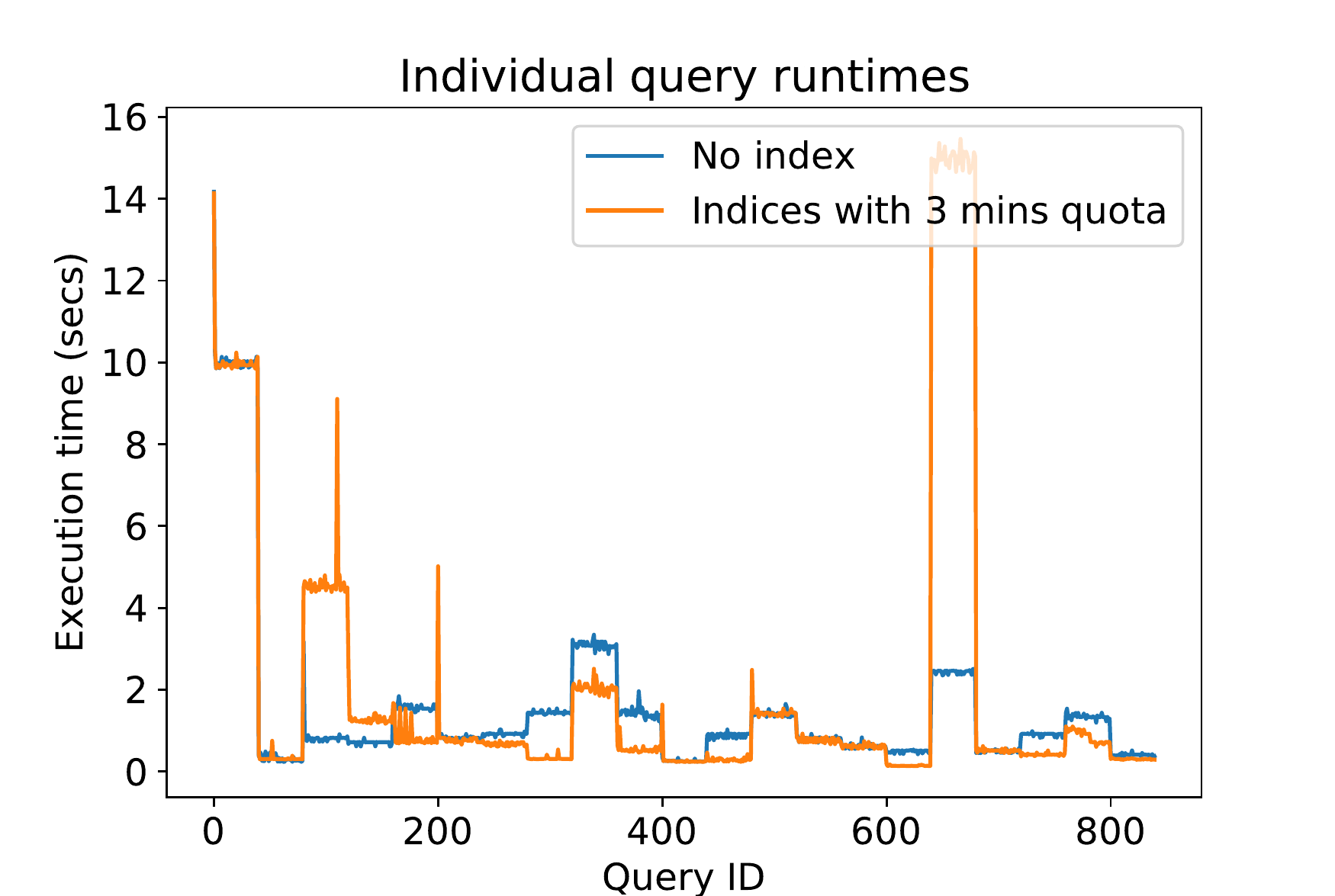}
  \caption{Runtime for each query in the workload under no indexes and under indexes recommended with a three-minute time budget.  Although most blocks of queries are faster with indexes as expected, in a few cases the optimizer chooses a bad plan and performance is significantly worse.}
  \label{fig:zerovs3min}
  \end{subfigure}
  \begin{subfigure}{0.45\textwidth}
    \includegraphics[width=1\columnwidth]{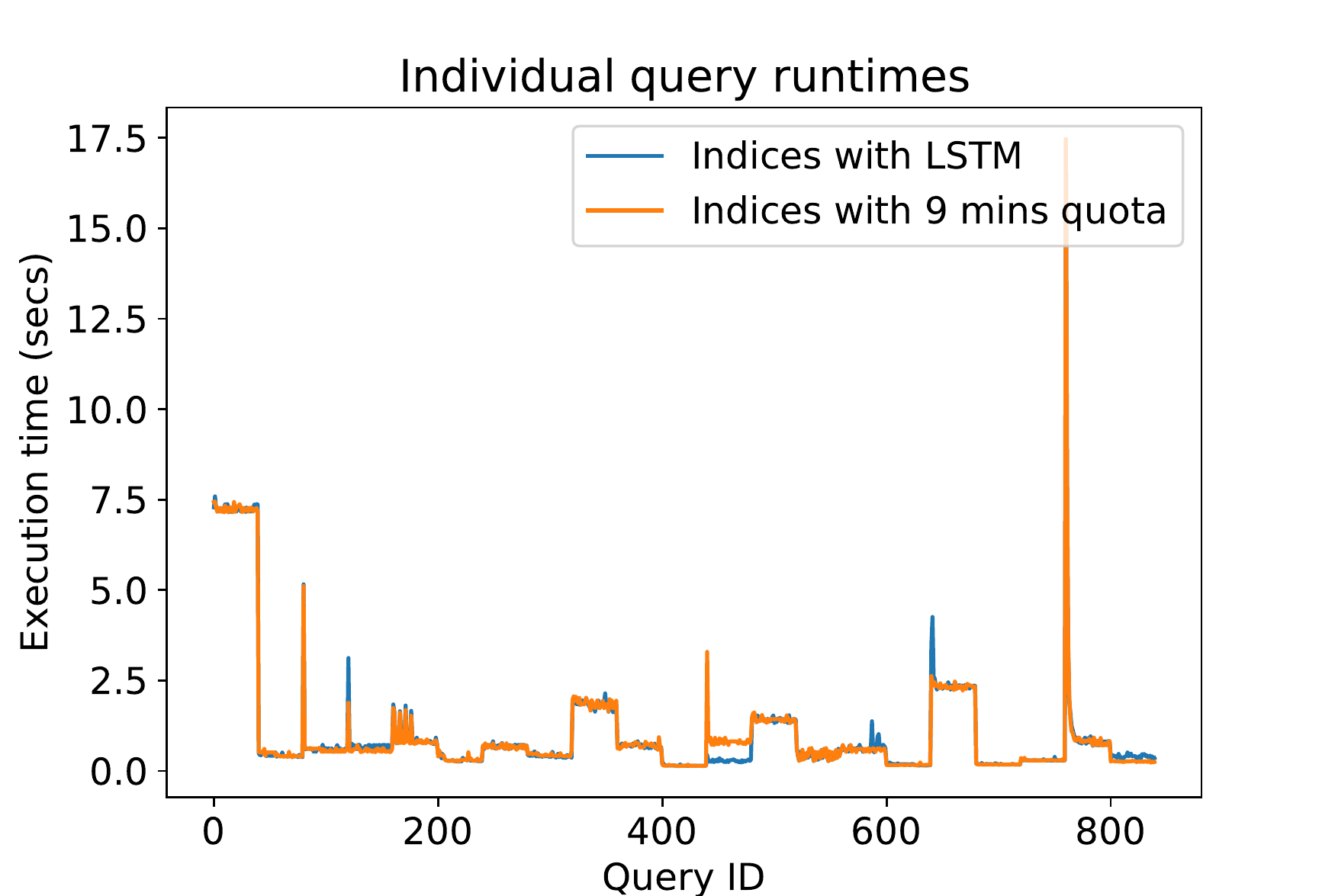}
    \caption{Runtime for each query in the TPC-H workload under our LSTM-based pre-compression scheme and the ``optimal'' indexes recommended by SQL Server.  The pre-compressed workload achieves essentially identical performance but with significantly smaller time budgets.}
    \label{fig:lstmvsoptimal}
  \end{subfigure}  
  \label{fig:queryruntime}
  \caption{Comparing runtimes for all queries in the workload under different index recommendations.}
\end{figure*}

Surprisingly, under tight time budgets, the index recommendations made by the native system can actually \emph{hurt} performance relative to having no indexes at all! The optimizer chooses a bad plan based on the suboptimal indexes.  In Figure \ref{fig:zerovs3min} and Figure \ref{fig:lstmvsoptimal}, we show the sequence of queries in the workload on the x-axis, and the runtime for each query on the y-axis. The indexes suggested under a 3 minute time budget result in all instance of TPC-H query 18 (queries 640-680 in Figure \ref{fig:zerovs3min}) taking much longer than they would take when run without these indexes. 

The key conclusion is that \emph{pre-compression can help achieve the best possible recommendations in significantly less time, even though compression is already being applied by the engine itself.} 

\begin{figure}
\centering
\includegraphics[keepaspectratio=true, width=\columnwidth]{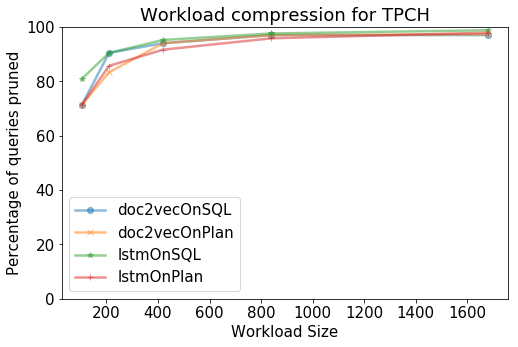}
\caption{Compression achieved on the TPC-H workloads at different scales. }
\label{tpch-compression}
\vspace{-1em}
\end{figure}

\paragraph*{Compression ratio}
Figure \ref{tpch-compression} shows the compression (percentage reduction in workload size) achieved by our algorithms on the TPC-H workload. We see that as the number of queries in the workload increases, the overall compression improves. This increase in compression this because our workloads at different were generated using 21 TPC-H query types. Since our algorithms looks for syntactic equivalences,  they result in a summary which is always around 20-30 queries in size.

\paragraph*{Time taken for summarization}
All of our methods, take about 2-3 minutes to summarize the workload. This excludes the offline one-time training step for learning the query2vec model (we discuss query2vec training time in Section \ref{sec:training_time}). The index suggestion task is roughly quadratic and takes more than an hour on the complete workload, so the time taken for summarization is amply justified.   


\subsection{Classifying Queries by Error}
\paragraph*{Classifying multiple errors}
In this experiment, we use the workload Snowflake-MultiError as described in Table \ref{eval_dataset_details} to train a classifier which can predict an error type (or no\_error) given an input query. We use the pre-trained LSTM autoencoder based Query2Vec model on the Snowflake dataset\footnote{The performance on Doc2Vec based model is similar.}. We evaluate only the models trained on query strings, because plans for some queries were not available since the corresponding database objects were deleted or the schema was altered. 

We use the datasets in (Table \ref{dataset_details}) to learn the query representations for all queries in the Snowflake-MultiError workload. Next, we randomly split the learned query vectors (and corresponding error codes) into training ($85\%$) and test ($15\%$) sets. We use the training set to train a classifier using the Scikit-learn implementation of Extremely Randomized Trees \cite{scikit-learn,sklearn_api}. We present the performance of this classifier on the test set. 

\begin{table}[h]
\centering
\begin{tabular}{|l|l|l|l|l|}
\hline
Error Code  & precision & recall & f1-score & \# queries \\ \hline
-1          & 0.986     & 0.992  & 0.989    & 7464       \\
604         & 0.878     & 0.927  & 0.902    & 1106       \\
606         & 0.929     & 0.578  & 0.712    & 45         \\
608         & 0.996     & 0.993  & 0.995    & 3119       \\
630         & 0.894     & 0.864  & 0.879    & 88         \\
2031        & 0.765     & 0.667  & 0.712    & 39         \\
90030       & 1.000     & 0.998  & 0.999    & 1529       \\
100035      & 1.000     & 0.710  & 0.830    & 31         \\
100037      & 1.000     & 0.417  & 0.588    & 12         \\
100038      & 0.981     & 0.968  & 0.975    & 1191       \\
100040      & 0.952     & 0.833  & 0.889    & 48         \\
100046      & 1.000     & 0.923  & 0.960    & 13         \\
100051      & 0.941     & 0.913  & 0.927    & 104        \\
100069      & 0.857     & 0.500  & 0.632    & 12         \\
100071      & 0.857     & 0.500  & 0.632    & 12         \\
100078      & 1.000     & 0.974  & 0.987    & 77         \\
100094      & 0.833     & 0.921  & 0.875    & 38         \\
100097      & 0.923     & 0.667  & 0.774    & 18         \\
\hline
avg / total & 0.979     & 0.979  & 0.978    & 15000     \\
\hline
\end{tabular}
\caption{Performance of classifier using query embeddings generated by LSTM based autoencoder for different error types (-1 signifies no error).}
\label{tab:classifyAll}
\end{table}

We summarize the performance of classifier on error classes with more than 10 queries each in Table \ref{tab:classifyAll}. The classifier performs well for the errors that occur sufficiently frequently, suggesting that the syntax alone can indicate queries that will generate errors.  This mechanism can be used in an online fashion to route queries to specific resources with monitoring and debugging enabled to diagnose the problem.  Offline, query error classification can be used for forensics; it is this use case that was our original motivation.  
Although individual bugs are not difficult to diagnose, there is a long tail of relatively rare errors; manual inspection and diagnosis of these cases is prohibitively expensive.  With automated classification, the patterns can be presented in bulk.

\paragraph*{Classifying out-of-memory errors}
In this experiment, we compare the classification performance of our method for one type of error considered a high priority for our colleagues --- queries running out of memory (OOM). We compare to a baseline heuristic method developed in collaboration with Snowflake based on their knowledge of problematic queries.  We use the workload Snowflake-OOM as described in Table \ref{eval_dataset_details} to train a classifier to predict out-of-memory errors. Following the methodology in the previous classification task, we use the pre-trained Query2Vec model to generate query representations for the workload, randomly split the learned query vectors into training ($85\%$) and test ($15\%$) sets, train a classifier using the Scikit-learn implementation of Extremely Randomized Trees, and present the performance on the test set.

\emph{Heuristic Baselines: } We interviewed our collaborators at Snowflake and learned that the presence of window functions or joins between large tables in the queries tend to be associated with OOM errors. We implement four na\"{i}ve baselines that looks for the presence of window functions or a join between at least 3 of the top $1000$ largest tables in Snowflake. The first baseline looks for the presence of heavy joins, the second baseline looks for window functions, and the third baseline looks for the presence of either one of the indicators: heavy joins \textbf{or} window functions, and the fourth baseline looks for the presence of both heavy joins \textbf{and} window functions.  The baselines predicts that the query will run out of memory if the corresponding indicator is present in the query text.


Table \ref{tab:classifyOOM} shows the results. We find that our method significantly outperforms the baseline heuristics, without requiring any domain knowledge or custom feature extractors. We do find that the presence of heavy joins and window functions in the queries are good indicators of OOM errors (specially if they occur together) given the precision of these baselines, however, the low recall suggests that such hard-coded heuristics would miss a other causes of OOM errors. Query2Vec obviates the need for such hard-coded heuristics. 
As with any errors, this mechanism can be used to route potentially problematic queries to clusters instrumented with debugging or monitoring harnesses, or potentially clusters with larger available main memories.  We see Query2Vec as a component of a comprehensive scheduling and workload management solution; these experiments show the potential of the approach.


\begin{table}[ht]
\centering
\begin{tabular}{|L{3.4cm}|l|l|l|l|}
\hline
Method  & precision & recall & f1-score  \\ \hline
Contains heavy joins          & 0.729     & 0.115  & 0.198       \\  \hline
Contains window funcs         & 0.762     & 0.377  & 0.504       \\  \hline
Contains heavy joins OR window funcs         & 0.724     & 0.403  & 0.518       \\  \hline
Contains heavy joins AND window funcs         & 0.931     & 0.162  & 0.162       \\  \hline
Query2Vec-LSTM         & \textbf{0.983}     &  \textbf{0.977}  & \textbf{0.980}       \\ \hline
Query2Vec-Doc2Vec         & 0.919     &  0.823  & 0.869       \\ 
\hline
\end{tabular}
\caption{Classifier performance predicting OOM errors for various baselines and Query2Vec based models. LSTM autoencoder based models significantly outperform all the other methods.}
\label{tab:classifyOOM}
\end{table}



\subsection{Training Time}

\begin{figure}[ht]
  \centering
  \begin{subfigure}{0.45\textwidth}
  \includegraphics[keepaspectratio=true, width=\columnwidth]{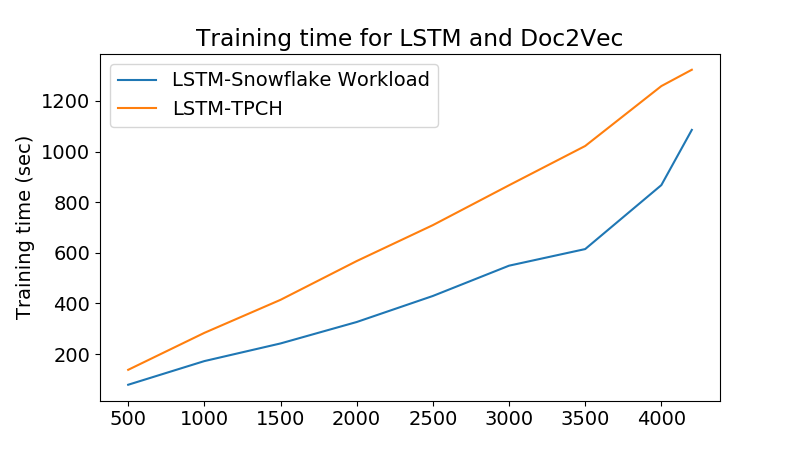}
  \end{subfigure}
  \begin{subfigure}{0.45\textwidth}
    \includegraphics[keepaspectratio=true, width=\columnwidth]{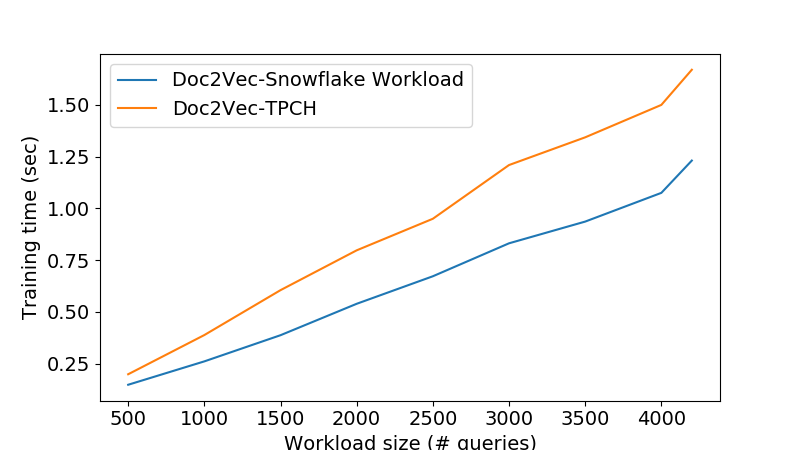}
  \end{subfigure}  
  \caption{Time taken for training the two Query2Vec models, the plots show the training time LSTM autoencoder and Doc2vec respectively. LSTM based methods take longer to train.}
  \label{fig:trainingtime}
\end{figure}

In this experiment, we evaluated the training time of the embedding model against datasets of varying sizes.  The hete
Figure \ref{fig:trainingtime} shows the trend for training times for learning the Query2Vec models on the two dataset we use (Table \ref{dataset_details}) for varying workload size. All of the models were trained on a personal computer (MacBook Pro with 2.8 GHz Intel Core i7 processor, 16 GB 1600 MHz DDR3 and NVIDIA GeForce GT 750M 2048 MB graphics card). We use the highly parallelized publicly available implementation of doc2vec \cite{doc2vecgensim}.  For small workloads, training time is a few seconds.  The entire Snowflake training set of $500,000$ queries takes less than 10 minutes.  

\label{sec:training_time}

We implemented the LSTM autoencoder in PyTorch, without any attempt to optimize the runtime. This method takes longer to train, going up to 20 minutes for workload sizes of 4200 queries. The total training time using the entire Snowflake workload was roughly 14 hours. We find that this time is justified, given the increase in performance and that this is onetime offline step. Furthermore, faster neural architectures can be deployed to greatly improve this time. We leave this optimization for future work.

We find that the training time is a function of average query length in the workload. More concretely, training time increases with the increase in average query length.

\subsection{Transfer Learning and Model Reuse}
\label{sec:transferlearning}

In this section we evaluate if the Query2Vec models trained on a large number of queries from one workload (or a combination of workloads) can be used to learn query representations for separate previously unseen workload. 
To this end, we measure the performance of workload summarization algorithm on TPC-H dataset, while using the Query2Vec model trained on Snowflake dataset (Table \ref{dataset_details}. The experimental workflow is as follows: 1) Train Query2Vec models on Snowflake workload, 2) Infer query representations for TPC-H workload, 3) Generate a workload summary, 4) Measure the performance of this summary against the results in Section \ref{sec:summary-results}.

\begin{figure}[ht]
\centering
\includegraphics[keepaspectratio=true, width=\columnwidth]{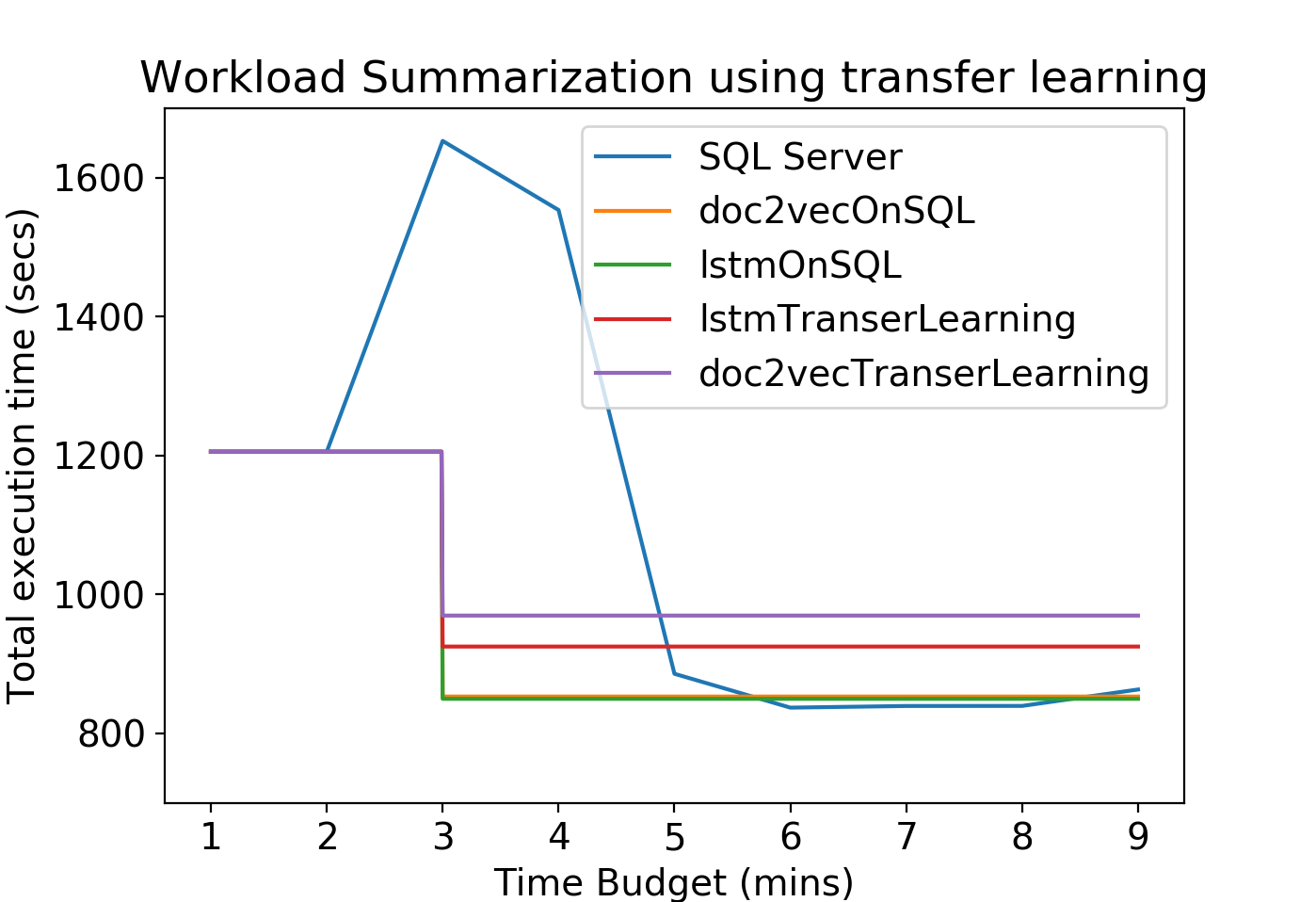}
\caption{Performance of workload summarization algorithm when using a pre-trained query2vec model for Snowflake workload to generate the representation for queries in the TPC-H workload. }
\label{fig:transferlearning}
\end{figure}

We summarize the results for this experiment in Figure \ref{fig:transferlearning}. We find that even with no prior knowledge of the TPC-H workload, query2vec model trained on Snowflake workload aids workload summarization and even outperforms the index recommendation by SQL Server till the time budget $<5$ minutes. This validates our hypothesis that \textit{we can train these generic querv2vec models on a large number of queries offline, and use these pre-trained models for workload analytics on unseen query datasets.} 

\section{Related Work}
\label{related_work}
\paragraph*{Word Embeddings and representation learning}
Word embeddings were introduced by Bengio et al. \cite{bengio2003neural} in 2003, however the idea of distributed representations for symbols is much older and was proposed earlier by Hinton in 1986 \cite{hinton1986learning}.
Mikolov et al. demonstrated \textit{word2vec} \cite{mikolov2013efficient,mikolov2013distributed}, an efficient algorithm which uses negative sampling to generate distributed representations for words in a corpus. \textit{word2vec} was later extended to finding representations for complete documents and sentences in a follow-up work, \textit{doc2vec} \cite{le2014distributed}.
Levy et al. provided a theoretical overview of why these techniques like \textit{word2vec} \cite{levy2014linguistic,goldberg2014word2vec,pennington2014glove} work so well. 
Generic methods for learning representations for complex structures were introduced in \cite{rudolph2016exponential, bengio2013representation}. In contrast, our work provided specialized algorithms for learning representations for SQL queries.
Representation learning for queries has been implicitly used by Iyer et el. \cite{srini_summarization} in some recent automated code summarization tasks, using a neural attention model, whereas we evaluate general query embedding strategies explicitly and explore a variety of tasks these embeddings enable.
Zamani et al. \cite{search_query2vec} and Grbovic et al. \cite{grbovic2015context} proposed a method to learn embeddings for natural language queries or to aid information retrieval tasks, however we consider learning embeddings for SQL queries and their applications in workload analytics.
LSTMs have been used various text encoding tasks like sentiment classification by Wang et el. \cite{wang2016attention}, machine translation by Luong et el. \cite{luong2015effective} and as text autoencoders by Li et el. \cite{lstm_autoencoders}. Our work is inspired by the success of these approaches and demonstrates their utility for SQL workload analytics.
  
\paragraph*{Compressing SQL Workloads}
Workload summarization has been done in the past by Surajit's et al. \cite{chaudhuri2002compressing}, however their method does not use query embeddings. Our evaluation strategy (index selection as a method to evaluate the compression) for workload summarization remains the same as the one proposed in \cite{chaudhuri2002compressing}.
Piotr Kolaczkowski et al. provided a similar solution to workload compression problem \cite{kolaczkowski2008compressing}, however their work doesn't provide a detailed evaluation or the dataset they perform the evaluation on, therefore we do not provide a comparison with their approach.


\paragraph*{Workload analytics and related tasks}
Decades ago, Yu et al. characterized relational workloads as a step toward designing benchmarks to evaluate alternative design tradeoffs for database systems \cite{yu:92}.
Jain et al. report on the SQL workload generated from a multi-year deployment of a database-as-a-service system \cite{shrjainSQLShare}, finding that the heterogeneity of queries was higher than for conventional workloads and identifying common data cleaning idioms \cite{jain:16a}.  We use this dataset in our study.
Grust et al. use a query workload to support SQL debugging strategies~\cite{grust:13} ; we envision that our embedding approaches could be used to identify patterns of mistakes and recommend fixes as a variant of our query recommendation task.
We envision workload analytics as a member of a broader class of services for weakly structured data sharing environments (i.e., ``data lakes"). Farid et al. propose mining for integrity constraints in weakly structured ``load first" environments (i.e., ``data lakes") \cite{farid:16}.  Although the authors do not assume access to a query workload, we consider the workload, if available, a source of candidate integrity constraints.
\balance

\section{Future Work}
\label{future_work}
We are interested in adapting this approach for other perennial database challenges, including query optimization and data integration.  Query embeddings can simplify the process of incorporating past experience into optimization decisions by avoiding an ever-expanding set of heuristics.  We also envision mapping SQL vectors to plan vectors automatically, short circuiting the optimization process when a workload is available.  Pre-trained Query2Vec models could assist in optimizing complex queries when the optimizer has incomplete information: in heterogeneous ``polystore'' environments~\cite{bigdawg} where an unproven system's performance may be unknown, or in real time situations where data statistics are not available. In these low-information situations, Query2Vec models can be used as another source of information to offset the lack of a robust cost model and complete statistics.  Looking further ahead, given enough (query, optimized plans) tuples, it could be possible to train a model to directly generate an optimized plan using an RNN in the same way a caption for an image can be synthesized!

Related to query optimization, we also intend to study query runtime prediction.  To the extent that runtime is a function of data statistics and the structure of the query, vector embeddings should be able to capture the patterns given a workload and enable predicting it.  The size of the workload required for training and the sensitivity to changes in the data are potential challenges. 

For data integration, we hypothesize that query embeddings can help uncover relationships between schema elements (attributes, tables) by considering how similar elements have been used in other workloads.  For example, if we watch how users manipulate a newly uploaded table, we may learn that ``la" corresponds to ``latitude" and ``lo" corresponds to ``longitude."  In weakly structured data sharing environments, there are thousands of tables, but many of them are used in the same way. Conventional data integration techniques consider the schema and the data values; we consider the query workload to be a novel source of information to guide integration.

We will also explore transfer learning further, i.e. training a Query2Vec model on a large shared public query corpus then reusing it in more specialized workloads and situations. We will also publish a pretrained query2vec model which database developers and researchers can utilize to assist with a variety of workload analytics tasks. 

In this paper we only considered embeddings for queries, but we intend to explore the potential for learning representations of the database elements as well: rows, columns, and complete tables. This approach could enable  new applications, such as automatically triggering relevant queries for never-before-seen datasets. That is, we could compare the vector representation for the newly ingested dataset with vector representations of queries to find suitable candidates. 

We are also exploring other neural architectures for query2vec, including TreeLSTMs \cite{tree_lstm} which can take query plan tree as an input in order to produce query embeddings, thus learning similarities not just in query text, but also in the plan trees.

Finally, wherever applicable, we will add these as features to existing database systems and services and conduct user studies to measure their effect on productivity.

\section{Conclusions}
\label{conclusion}
We presented Query2Vec, a general family of techniques for using NLP methods for generalized workload analytics.  This approach captures the structural patterns present in the query workload automatically, largely eliminating the need for the specialized syntactic feature engineering that has motivated a number of papers in the literature.

To evaluate this approach, we derived new algorithms for two classical database tasks: workload summarization, which aims to find a representative sample of a workload to inform index selection, and error forensics, which aims to assist DBAs and database developers in debugging errors in queries and to assist users when authoring queries by predicting that query might result in an error. 
On these tasks, the general framework outperformed or was competitive with previous approaches that required specialized feature engineering, and also admitted simpler classification algorithms because the inputs are numeric vectors with well-behaved algebraic properties rather than result of arbitrary user-defined functions for which few properties can be assumed. 
We find that even with out-of-the-box representation learning algorithms like Doc2Vec, we can generate query embeddings and enable various new machine learning based workload analytics tasks. To further improve on the accuracy of these analytics tasks, we can use more sophisticated training methods like LSTM based autoencoders. 

Finally, we demonstrate that learnings from models on large SQL workloads can be transfered to other workloads and applications.

We believe that this approach provides a new foundation for a variety of database administration and user productivity tasks, and will provide a mechanism by which to automatically adapt database operations to a specific query workload.
\section*{Acknowledgements}
We would like to thank our collaborators at Snowflake for valuable feedback and datasets which made this work possible. We would also like to thank Louis M Burger and Doug Brown from Teradata for their feedback and helpful discussion on the topics covered in this paper.  This work is sponsored by the National Science Foundation award 1740996 and Microsoft.
\balance

\bibliographystyle{abbrv}
\bibliography{references}
\end{document}